\documentclass[preprint2]{aastex}

\def\be{\begin{equation}}
\def\ee{\end{equation}}

\begin{document}

\title{New Electrodynamics of Pulsars} 

\author{Andrei Gruzinov}

\affil{CCPP, Physics Department, New York University, 4 Washington Place, New York, NY 10003}

\begin{abstract}

~~

We have recently proposed that Force-Free Electrodynamics (FFE) does not apply to pulsars -- pulsars should be described by the high-conductivity limit of Strong-Field Electrodynamics (SFE), which predicts an order-unity damping of the Poynting flux, while FFE postulates  zero damping. The strong damping result has not been accepted by several pulsar experts, who claim that FFE basically works and the Poynting flux damping can be arbitrarily small.  

Here we consider a thought experiment -- cylindrical periodic pulsar. We show that FFE is incapable of describing this object, while SFE predictions are physically plausible. The intrinsic breakdown of FFE should mean that the FFE description of the singular current layer (the only region of magnetosphere where FFE and the high-conductivity SFE differ) is incorrect. Then the high-conductivity SFE should be the right theory for real pulsars too, and the pure-FFE description of pulsars should be discarded.

~~

\end{abstract}

\section{Introduction} 

We have shown that ideal pulsars calculated in the high-conductivity limit of Strong-Field Electrodynamics (SFE
\footnote{Maxwell plus ${\bf j}=\sigma {\bf E}$ in the right frame (Gruzinov 2011a).}) dissipate an order-unity fraction of the Poynting flux in the singular current layer (SL), which in SFE exists only outside the light cylinder (Gruzinov 2011ab). This result, if true, is obviously important for interpreting pulsar phenomenology. SL should be the most powerful site of pulsar emission (cf Bai \& Spitkovsky 2010).

Two groups of prominent pulsar experts disagree with the strong-damping result (Li et al 2011, Kalapotharakos et al 2011) and claim that FFE (as it applies to pulsars) works exactly as it has always been thought to work, so that the SL damping can be arbitrarily small.

\begin{figure}
\plotone{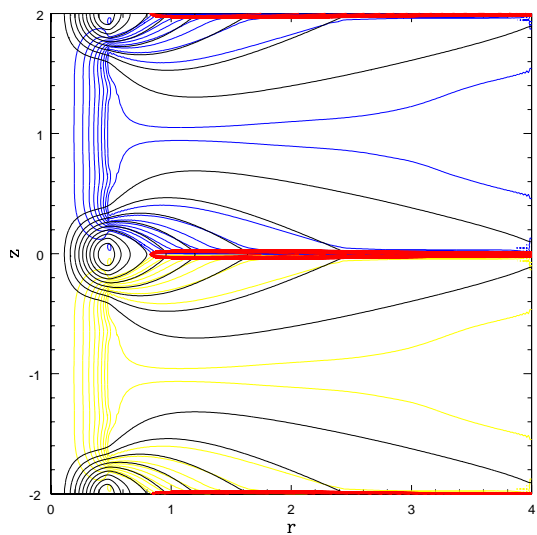}
\caption{The SFE magnetosphere. Star radius $r_s=0.5$, light cylinder radius $r=1$. Full period in $z$ is shown. {\it Thin yellow and blue:} isolines of integrated poloidal current of different signs. Note the current closing within the star. {\it Thicker black:} magnetic surfaces. {\it Very thick red:} positive isolines of $E^2-B^2$.} 
\end{figure}

Here we calculate FFE and SFE magnetospheres of an artificial system -- cylindrical periodic pulsar. We consider an ideally conducting cylinder rotating around its axis. The cylinder is magnetized axisymmetrically and periodically along the axis. 

It is clear without any calculations that the FFE magnetosphere of the cylindrical periodic pulsar is unphysical. We calculate the FFE magnetosphere anyway (\S2), not only as a counter for the SFE magnetosphere (\S3), but also to stress that FFE predictions are ambiguous and incorrect. The SFE magnetosphere, on the other hand, does look reasonable (Fig.1). 

The order of magnitude of the SL damping must be decided by the microphysics rather than by the global geometry of the problem. We therefore propose that SFE is in the right also when applied to real pulsars.

\section{FFE magnetosphere} 

The FFE magnetosphere is calculated by the standard CKF procedure (Contopoulos, Kazanas \& Fendt 1999): 
\be \label{pme}
(1-r^2)\Delta \psi-{2\over r}\partial _r\psi +F(\psi )=0,
\ee
\be 
\psi (r_s,z)=f(z),
\ee 
\be 
\psi (r>1,0)=\psi (1,0),
\ee 
\be 
\psi (r,H)=0.
\ee 
Here $r$ is the cylindrical radius,  the light cylinder is at $r=1$, $r_s$ is the radius of the cylindrical star, $H$ is the quarter-period, $\psi$ is the magnetic stream function and electric potential, $F\equiv A(dA/d\psi )$, $A$ is twice the integrated poloidal current.

The function $f(z)$ represents the surface magnetization of the star. For no particular reason, we set $r_s=0.5$, $H=1$, and 
\be 
f(z) \propto \sum _k (-1)^ke^{-20(z-2k)^2}.
\ee 
This gives well-isolated periodically repeating regions of alternating sign $\psi$. If the star were not rotating, each region of sign-definite $\psi$ would have field lines closing onto itself. 

In truth, there is  one ill-defined reason for choosing to have well-isolated regions of sign-definite $\psi$. One would think that these regions work roughly as separate pulsars, and the CKF procedure must be applicable to each of the ``pulsars'' in a more or less unmodified form. Then FFE is expected to make unambiguous predictions regarding the cylindrical pulsar and its SL.

\begin{figure}
\plotone{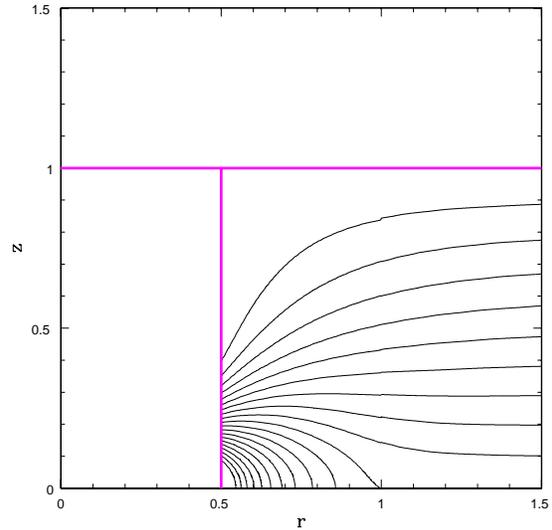}
\caption{The ``standard'' FFE magnetosphere. {\it Thick magenta:} boundary of the star and the quarter-period. {\it Thin black:} magnetic surfaces. The poloidal current closes within each half-period of sign-definite $\psi$ by flowing in the equatorial plane from infinity to the light cylinder and then flowing to the star in the magnetic separatrix (the last closed magnetic surface).} 
\end{figure}

\begin{figure}
\plotone{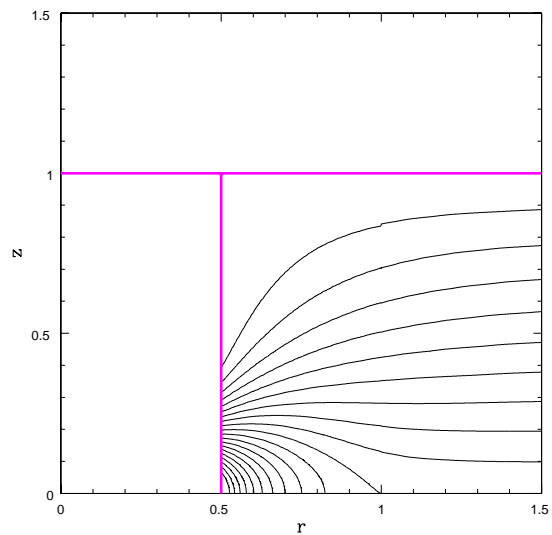}
\caption{A possible FFE magnetosphere. There is no singular poloidal current in the equatorial planes and in the magnetic separatrices. The current closes within the star.} 
\end{figure}

We first solve the problem (\ref{pme}) just as described by CKF, and get the quarter-period shown in Fig.2. Next we note that for our periodic pulsar the net poloidal current is always zero by symmetry, and we are actually free to choose any value for the SL poloidal current. For example, Fig.3 shows the FFE magnetosphere with zero poloidal current in the SL\footnote{ The equatorial plane outside the light cylinder remains singular, as it still carries a non-zero surface toroidal current and a non-zero surface charge. The only apparent difference between the two FFE magnetospheres is the usual puffing up of the magnetic separatrix in Fig.2 as compared to Fig.3. This is the effect of the separatrix current. The effect is rather weak because the separatrix current is small for our choice of parameters.}.

The ambiguity of the FFE magnetosphere is worrisome. Much worse is that all FFE magnetospheres are actually unphysical, because no matter how big the postulated SL poloidal current, the electromagnetic field inevitably becomes electric-like\footnote{ Meaning $E^2-B^2>0$. Electric-like electromagnetic field is not described by FFE.} at large enough distances from the light cylinder.  

Indeed, since the FFE magnetic surfaces are also the equipotentials, the electric field at large distances from the light cylinder depends only on $z$, while both poloidal and toroidal components of the magnetic field decrease as $r^{-1}$ at large $r$.

For normal pulsars, isolated and spherical, as calculated by FFE or the high-conductivity SFE, the open magnetic surfaces diverge from each other at large $r$, allowing the electric field to decrease in such a way that the electromagnetic field apparently remains everywhere magnetic-like, except in the equator outside the light cylinder. But for our periodic pulsar, the neighboring magnetic surfaces cannot diverge to more than a half-period.

\section{SFE magnetosphere} 

The SFE magnetosphere is calculated as described by Gruzinov (2011a), only now we use periodic vertical boundary conditions, change the shape of the star, and  choose the external toroidal current flowing within the star in such a way as to  roughly reproduce the characteristic width of the sign-definite regions of $\psi$ on the stellar surface. The parameters of the numerical simulation (in terms of the angular velocity of the star $\Omega$, with $c=1$ and with $\Omega =1$ in Fig.1) are
\begin{itemize}

\item{radius of the star $r_s=0.5\Omega ^{-1}$}

\item{electric conductivity of the star $\sigma _s=400\Omega$}

\item{electric conductivity of vacuum $\sigma =40\Omega$}

\item{regularizing diffusivity $\eta=0.0003\Omega ^{-1}$}

\end{itemize}

Fig.1 shows that now, in the SFE magnetosphere, the electromagnetic field  becomes electric-like only in the SLs (planes $z=2k$ outside the light cylinder). The entire Poynting flux is now damped\footnote{Poynting flux damping is about 50\% for a normal axisymmetric pulsar (Gruzinov 2011b).}, thus allowing the magnetic surfaces (equipotentials) to close, and the electric field to decrease. Just like for isolated pulsars, the solution is given by ``FFE with correct boundary conditions at SL'' (Gruzinov 2011b).

\section{Conclusions} 

Large Poynting flux damping is inevitable for cylindrical pulsars. It seems unlikely that the very existence of the local SL damping can be decided by the global geometry. Therefore, SFE rather than pure FFE  must be in the right for real pulsars. 

The final word, obviously, belongs to microphysics, which is anyway needed to calculate the pulsar emission. If SFE is right, the microphysics handles an order-unity fraction of energy, at least right beyond the light cylinder.

\acknowledgements

I thank Anatoly Spitkovsky for a valuable bit of information.

\end{document}